\documentclass[10pt,conference]{IEEEtran}
\IEEEoverridecommandlockouts
\usepackage{cite}
\usepackage{amsmath,amssymb,amsfonts}
\usepackage{algorithmic}
\usepackage{graphicx}
\usepackage{textcomp}
\usepackage{xcolor}
\usepackage{multicol, blindtext}
\def\BibTeX{{\rm B\kern-.05em{\sc i\kern-.025em b}\kern-.08em
    T\kern-.1667em\lower.7ex\hbox{E}\kern-.125emX}}

\usepackage{array}
\newcolumntype{L}[1]{>{\raggedright\let\newline\\\arraybackslash\hspace{0pt}}m{#1}}
\newcolumntype{C}[1]{>{\centering\let\newline\\\arraybackslash\hspace{0pt}}m{#1}}
\newcolumntype{R}[1]{>{\raggedleft\let\newline\\\arraybackslash\hspace{0pt}}m{#1}}
\newcommand{\NA}{---}

\usepackage{booktabs}
\newcommand{\ra}[1]{\renewcommand{\arraystretch}{#1}}

\IEEEoverridecommandlockouts\IEEEpubid{\makebox[\columnwidth]{978-3-903176-15-7~\copyright~2019 IFIP \hfill} \hspace{\columnsep}\makebox[\columnwidth]{ }}
\begin{document}
\title{On the classification and false alarm of invalid prefixes in RPKI based BGP route origin validation
}

  \author{\IEEEauthorblockN{Wenjie Xu}
  \IEEEauthorblockA{\textit{Department of Electronic Engineering} \\
  \textit{Tsinghua University}\\
  Beijing, China \\
  jackxu199603@gmail.com}
  \and
  \IEEEauthorblockN{Deliang Chang}
  \IEEEauthorblockA{\textit{Department of Electronic Engineering} \\
  \textit{Tsinghua University}\\
  Beijing, China \\
  chdlgs@gmail.com}
  \and
  \IEEEauthorblockN{Xing Li}
  \IEEEauthorblockA{\textit{Department of Electronic Engineering} \\
  \textit{Tsinghua University}\\
  Beijing, China \\
  xing@cernet.edu.cn}
}

\maketitle

\begin{abstract}
BGP is the default inter-domain routing protocol in today's Internet, but has serious security vulnerabilities\cite{murphy2005bgp}. One of them is (sub)prefix hijacking. IETF standardizes RPKI to validate the AS origin but RPKI has a lot of problems\cite{heilman2014consent}\cite{cooper2013risk}\cite{gilad2017we}\cite{gilad2017maxlength}, among which is potential false alarm. Although some previous work\cite{gilad2017we}\cite{heilman2014consent} points it out explicitly or implicitly, further measurement and analysis remain to be done.
Our work measures and analyzes the invalid prefixes systematically. We first classify the invalid prefixes into six different types and then analyze their stability. We show that a large proportion of the invalid prefixes very likely result from traffic engineering, IP address transfer and failing to aggregate rather than real hijackings.   
\end{abstract}


\section{Introduction}
Internet is an inter-connected network without a center. It is made up of more than 50000 autonomous systems(AS for short). To transmit packets across the ASes, \textbf{B}order \textbf{G}ateway \textbf{P}rotocol(BGP for short) is designed and implemented\cite{rekhter2005border}. In BGP, every autonomous system will announce the prefixes owned by itself and propagate routing information it learned from its neighbors according to a policy. When propagating the prefix, the ASes will maintain a path to the origin of the prefix and can choose among different paths. 
\par
However, as a fundamental part of Internet infrastructure, BGP has serious security vulnerabilities\cite{murphy2005bgp}. One of them is BGP prefix hijacking. In prefix hijacking, an AS may illegitimately announce a prefix not owned by itself and then those ASes who accept the announcement will transmit the packets to a wrong destination. Prefix hijacking may result from misconfiguration and malicious attack. Using prefix hijacking, the attackers can block web service, steal secret information and do man in the middle attack, etc\cite{murphy2005bgp}. Actually, BGP prefix hijacking is frequently appearing in recent news\cite{www.bishopfox.com}\cite{www.internetsociety.org}. In one of the most famous prefix hijacking events, Pakistan Telecom blocked Youtube for more than 2 hours\cite{hijacking2008ripe}, causing inconvenience to the Youtube visitors all over the world. To tackle the problem of prefix hijacking, IETF(Internet Engineering Task Force) standardizes a framework called RPKI(Resource Public Key Infrastructure) to validate the origination AS of a prefix\cite{mohapatra2013bgp}. In RPKI, trust anchors(the five regional Internet registries) sign prefixes and allocate the signed prefixes to NIRs, LIRs or ISPs. And then prefixes will be assigned hierarchically to customers. Prefix owners can sign an object called \textbf{R}oute \textbf{O}rigin \textbf{A}uthorization to authorize an AS to announce a prefix. A ROA consists of a prefix, prefix length, maximum length, the AS authorized to announce the prefix\cite{mohapatra2013bgp} and the trust anchor. ROAs are stored in distributed repositories and can be fetched to validate the BGP items. We now illustrate how ROAs can be used to validate the BGP items. For simplicity, we define BGP item as a two-element tuple (prefix, AS path), where AS path means a sequence of ASes that the announcement of the prefix traversed. And the last AS in the AS path is the origination AS of the prefix. Given a set of ROAs, there are three possible validation results of a BGP item as shown below.
\begin{description}
\item[\textbf{Unknown}]\qquad The prefix in BGP item is not covered(Prefix A covering B means B is not longer then A and the first length of prefix A bits of the two prefixes coincide.) by any prefix in ROA.
\item[\textbf{Valid}]\qquad There exists a ROA item such that the prefix in BGP item is covered by the prefix in ROA, the length of the prefix in BGP item is no longer then the maximum length and the origination AS of the BGP item is the same as the authorized AS in ROA.
\item[\textbf{Invalid}]\qquad The prefix in the BGP item is covered by one prefix in ROA, but is not valid.
\end{description}
\par 
With ROAs validating all the BGP items, the prefix hijacking problem seems to be solved perfectly. However, up to now, RPKI has not been fully deployed and partial deployment may result in unexpected trouble like false alarm. False alarm prefix can reduce the trustability of RPKI and even make those ASes that discard false invalid prefixes lose Internet connection to the discarded prefixes. So a systematic analysis and evaluation of current invalid BGP prefixes are in urgent needs and of great significance. Our work, to the best of our knowledge, systematically classifies and evaluates the invalid BGP prefixes for the first time. We find that most of the invalid prefixes result from traffic engineering purposes like multi-homing and load-balancing. We also find a large part of the invalid prefixes are very likely transfer prefixes. And finally we build a website to publish our analysis and classification result to help the network operators design better routing policy.         
\section{Related work} 
Currently there are mainly two lines of efforts to tackle the threat of prefix hijacking. The first is detection approach. For example, Zheng Zhang, et al., designed a system called iSPY to detect IP prefix hijacking on its own\cite{zhang2008ispy}. Xiaoliang Zhao, et al., analyzed BGP multiple origin conflicts and gave the potential reasons\cite{zhao2001analysis}. Xingang Shi, et al., utilized the correlation of control plane information and data plane information to detect the hijacking prefix\cite{shi2012detecting}. However, detection approach may suffer from false alarm. 
\par
The second is validation approach. To tackle the threat of prefix hijacking thoroughly, IETF(Internet Engineering Task Force) standardizes a framework called RPKI\cite{mohapatra2013bgp}. ASes can use signed ROAs to validate the origin of prefixes. However, RPKI is not perfect and faces some unexpected problems in the deployment process. Danny Cooper, et al., flipped the threat model and analyzed the risk of misbehaving RPKI authorities\cite{cooper2013risk}. Ethan Heilman, et al., design tools to detect potential harm to BGP prefix and propose some modification of RPKI to improve its transparency\cite{heilman2014consent}. Yossi Gilad, et al., show that MaxLength can be harmful to RPKI\cite{gilad2017maxlength}. Yossi Gilad, et al., point out that partial deployment of RPKI can result in false invalid BGP prefix\cite{gilad2017we}. However, to the best of our knowledge, there is no systematic measurement and analysis of invalid prefix. To fill the gap, we collect BGP prefixes for 3 months and analyze them by classifying them according to the AS path structure, aggregation structure and AS commercial relationship.



\section{Classifications of Invalid Prefixes}
\label{sec:ClaInP}
Now we describe our classifications of invalid prefixes based on AS-path structure, prefix aggregation structure and AS-relationship structure. The following six types of invalid prefixes are in fact false alarms, which mean invalid BGP items providing legitimate connections. Note that in all the invalid prefix illustration figures the AS following the prefix is the BGP origin AS and TAxx represents a trust anchor. 
\subsection{Invalid load-balancing prefix} 
\begin{figure}[htbp]
\includegraphics[width=0.45\textwidth]{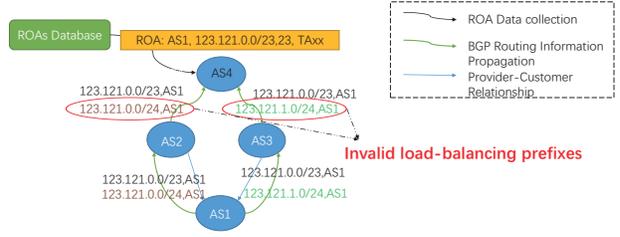}
\caption{An illustration of invalid load-balancing prefix}
\label{fig:invloadbal}
\end{figure}

In this scenario, an AS may first got a ROA with a relatively short maximum length. However, the AS may then announce to be the origin of a more specific prefix for load balancing reasons. For illustration, as shown in figure \ref{fig:invloadbal}, AS1 may be assigned a prefix 123.121.0.0/23 and to secure the prefix, relevant authority may sign a ROA with maximum length 23 for the (AS1, 123.121.0.0/23) pair. However, after some time, AS1 may announce two more specific prefixes 123.121.0.0/24 and 123.121.1.0/24 through its two providers AS2 and AS3 respectively for load balancing reason. Then, AS4 using RPKI for BGP prefix validation will determine those two more specific prefixes are invalid. 


\subsection{Invalid failing to aggregate prefix}
\begin{figure}[htbp]
\includegraphics[width=0.45\textwidth]{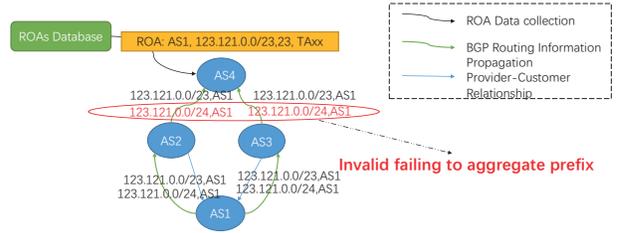}
\caption{An illustration of invalid failing to aggregate prefix}
\label{fig:failing2aggr}
\end{figure}

This type of invalid prefixes differ from invalid load-balancing prefixes in that invalid failing to aggregate prefixes are announced according to exactly the same export policy. For illustration, as shown in figure \ref{fig:failing2aggr}, AS1 may be first assigned a prefix 123.121.0.0/23 and a corresponding ROA with maximum length 23. Then, for some reason such as ignorance or customer requirement, AS1 may then announced a more specific prefix 123.121.0.0/24. However, because 24 is larger than the maximum length 23 in ROA, 123.121.0.0/24 is considered invalid by AS4, which uses RPKI for prefix validation.          
\subsection{Invalid multihoming prefix}
\begin{figure}[htbp]
\includegraphics[width=0.45\textwidth]{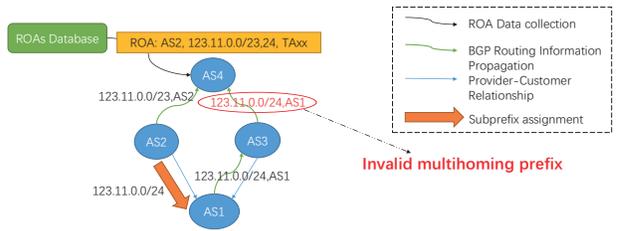}
\caption{An illustration of invalid multihoming prefix}
\label{fig:invmultihome}
\end{figure}

In this scenario, a customer doing multihoming may announce invalid prefix. For illustration, as shown in figure \ref{fig:invmultihome}, the provider AS2 may assign a subprefix 123.11.0.0/24 of its own prefix 123.11.0.0/23 to AS1 and at the same time, AS2 got a ROA item: AS2, 123.11.0.0/23,24, TAxx. However, AS1 may have other provider AS3 besides AS2 and propagate prefix thorough AS3. Then AS4 will treat 123.11.0.0/24 as invalid prefix. 

\subsection{Invalid singlehoming prefix}
\begin{figure}
\includegraphics[width=0.45\textwidth]{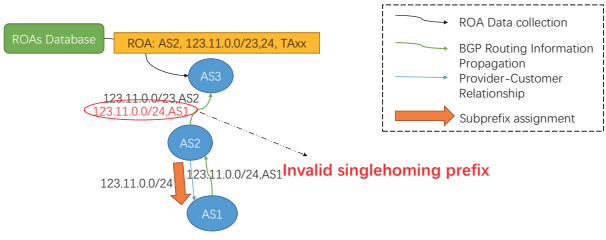}
\caption{An illustration of invalid singlehoming prefix}
\label{fig:invsinglehome}
\end{figure}

In this scenario, a customer doing singlehoming may also announce invalid prefix. For illustration, as shown in figure \ref{fig:invsinglehome}, AS2 has a prefix 123.11.0.0/23 and to protect its prefix from being hijacked,  AS2 also gets a ROA item: AS2, 123.11.0.0/23, 24, TAxx. However, AS2 further assigns a subprefix of 123.11.0.0/23 to its customer AS1. And AS1  announces the subprefix through its provider AS2. For some reasons like ignorance or to attract more traffic, AS2 does not aggregate the subprefix announced by the customer AS1. Then AS3 considers 123.11.0.0/24 as invalid.  

\subsection{Invalid provider prefix}
\begin{figure}[htbp]
\includegraphics[width=0.45\textwidth]{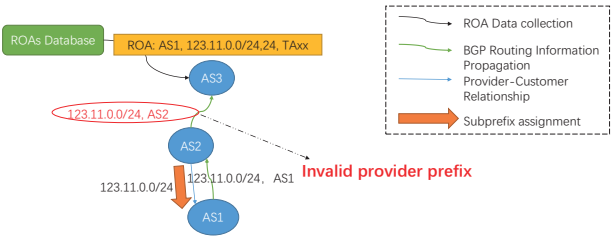}
\caption{An illustration of invalid provider prefix}
\label{fig:invpro}
\end{figure}

A dual case of invalid singlehoming prefix is invalid provider prefix. Actually, some of the providers do not include its customer's AS number into the AS path when propagating the prefix announced by its customer. For illustration, as shown in \ref{fig:invpro}, the provider AS2 assigns a prefix 123.11.0.0/24 to its customer AS1. To secure the prefix 123.11.0.0, a ROA (AS1, 123.111.0.0/24, 24, TAxx) is signed and published in the ROA database. However, when propagating the routing information, AS2 announces 123.111.0.0/24 as the origin AS though it should have included AS1 as the origin AS. As a result, 123.11.0.0/24 will be considered as an invalid prefix.      
\subsection{Invalid transfer prefix}
\begin{figure}
\includegraphics[width=0.45\textwidth]{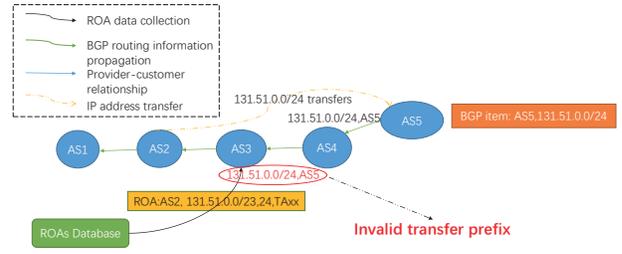}
\caption{An illustration of invalid transfer prefix}
\label{fig:invtrans}
\end{figure}

Last but not least, due to active IP address transaction and the mobility of the organizations owning IP prefixes, IP address transfer is becoming more and more frequent and can result in invalid prefix. For illustration, as shown in figure \ref{fig:invtrans}, AS2 used to own the prefix 131.51.0.0/24 and the corresponding ROA. For some reason, the prefix transferred to AS5 but the ROA: (AS2, 131.51.0.0/23, 24, TAxx) isn't revoked or modified. So when the routing information of 131.51.0.0/24 is propagated to AS3 from AS5, the prefix is considered invalid due to the obsolete ROA item.      

\section{Dataset and Route Origin Validation Result}
The six types of invalid prefixes described in section \ref{sec:ClaInP} are all actually \textbf{legitimate}. That is to say, they are \textbf{false invalid} prefixes. To evaluate different types of invalid prefixes, we first collect BGP routing data for almost 3 months from February, 2018 and then do route origin validation. To collect BGP routing data, we set up a private AS. Then with our private AS, we do BGP peering with AS4538(China Education and Research Network), which is adjacent to several core ASes and can collect most of the BGP data on today's Internet. Through AS4538, we collect BGP routing table, taking a snapshot of it every day, and BGP update data of the whole Internet. As for the ROA data, we use the software rpki-validator\cite{rpki-validator} provided by RIPE-NCC to collect and validate the ROA data. We also take everyday snapshot of the ROA data. 
After data collection, we build a prefix aggregation forest according to the aggregation relationship. We call the prefixes not covered by other prefixes maximal prefixes. Then with BGP routing table and ROA data, we do route origin validation. The validation result is shown in table \ref{tab:ValRes}. 
\begin{table}
\ra{1.3}
\centering 
\begin{tabular}{@{}lll@{}}
\toprule
Validation Result&Number of Routing Items&Ratio\\
\midrule
Unknown&635412&90.87\% \\
Valid&58931&8.43\%\\
Invalid&4949&0.71\%\\
\bottomrule 
\end{tabular}
\newline 
\caption{RPKI based Route Origin Validation Result(Data collected on 16th, May, 2018)}
\label{tab:ValRes}
\end{table}

\section{Are invalid prefixes really invalid?}
Theoretical analysis in section \ref{sec:ClaInP} shows that there are six possible scenarios where traffic engineering, prefix deaggregation and address transfer can result in \textbf{false invalid} prefixes. And notably, there are 4949 invalid prefixes in the validation result. Although address hijacking is frequent these days,  very unlikely we can detect thousands of hijackings at the same time. So the validation result is highly suspicious. 
\subsection{The classification result of real world BGP data}
According to the classifications listed in section \ref{sec:ClaInP}, we design classification rules as shown in table \ref{tab:ClaRule} to classify the BGP prefixes.

\begin{table*}[htbp]
\ra{1.3}
\centering 
\begin{tabular}{@{}L{0.13\textwidth}L{0.13\textwidth}L{0.13\textwidth}L{0.13\textwidth}L{0.13\textwidth}L{0.13\textwidth}L{0.13\textwidth}@{}}
\toprule
Type of invalid prefix&Is the AS in ROA the same as BGP origin AS&Is the AS in ROA provider of BGP origin AS& Is BGP origin AS  the provider of the AS in ROA&Multiple providers&Is there parent prefix or sibling prefix with different AS path&Is there parent prefix or sibling prefix with the same AS path\\
\midrule
Invalid load-balancing prefix& Yes      & No    &    No   &    \NA   &  Yes  & \NA            \\
Invalid failing to aggregate prefix& Yes      &No  &  No    &\NA           &   No    &Yes      \\
Invalid multihoming prefix&    No        &  Yes &    No   & Yes          &  Yes     & \NA     \\
Invalid singlehoming prefix&    No      &  Yes  &   No    &  No         &  Yes  & \NA   \\
Invalid provider prefix   &     No     &   No    & Yes   &  \NA         &  \NA    & \NA  \\
Invalid transfer prefix   &     No     &  No    & No      &  \NA         & \NA       &  \NA     \\
\bottomrule 
\end{tabular}
\newline 
\caption{The classification rules of invalid prefixes} 
\label{tab:ClaRule}
\end{table*}
We first sort the BGP prefixes, search the maximal prefix and build the prefix aggregation forest. For every node in the prefix aggregation forest, we associate it with an AS path attribute for search in classification process. Then we apply the rules in table \ref{tab:ClaRule} to efficiently classify the prefixes. To find transfer prefix, we also use zmap\cite{durumeric2013zmap} to scan the ip addresses under a prefix seeming to be transferred and we classify it as transfer prefix if we get active response. We show the classification result in table \ref{tab:ClaRes}. We observe that more than 60\% of the invalid prefixes very likely result from traffic engineering, IP address transfer and failing to aggregate rather than real hijacking. And the rest of the invalid prefixes can be other types of false invalid prefixes or real hijacking.       

\begin{table}[htbp]
\ra{1.3}
\centering 
\begin{tabular}{@{}L{0.08\textwidth}lL{0.08\textwidth}L{0.08\textwidth}L{0.08\textwidth}@{}}
\toprule
Type of Invalid prefix&Number&Percentage in invalid prefix&Number of long-lived (invalid prefix, origin AS) pairs&Percentage of prefixes with long-lived (prefix, origin AS) pair in this type\\
\midrule
Invalid load-balancing prefix&923&18.7\% &770  &83.4\% \\
Invalid failing to aggregate prefix&703&14.2\% &684 &97.3\%   \\
Invalid multihoming prefix&378&7.6\% & 355&93.9\% \\
Invalid singlehoming prefix&204&4.1\% &  177    & 86.8\%   \\
Invalid provider prefix&186&3.8\% & 147 & 79.0\% \\
Invalid transfer prefix& 737& 14.9\% & 658&  89.3\% \\
Other invalid prefix& 1818& 36.7\% & 1695 & 93.2\% \\ 
\bottomrule 
\end{tabular}
\newline 
\caption{The classification result and stability of invalid prefix(Data collected on May, 16th, 2018)}
\label{tab:ClaRes}
\end{table}

\subsection{The stability of invalid prefixes}
We also monitor the invalid prefixes from 28th, February, 2018 to 16th, May, 2018 and we find that as shown in table \ref{tab:ClaRes}, most of the invalid (prefix, origin AS)s in different types are actually long-lived(meaning the (prefix, origin AS) pair keeps existing during our data collection period), implying they are more likely just false alarms since the real hijackings tend to be short-lived. On the one hand, the potential false alarms diminish the reliability of RPKI, thus slowing down the deployment of RPKI. On the other hand, false alarm may affect the false invalid prefixes' reachability for RPKI adopters. So we build a website to publish the possible false alarm prefixes and help network operators make better routing policy to avoid losing reachability due to false alarm. The website can be accessed through 202.38.101.13:5000.    





\section{Conclusion and future work}
We first describe our classifications of invalid prefixes based on the AS path structure, prefix aggregation structure, AS commercial relationship. Then we collect real world BGP data for about 3 months and design rules to classify them. We find that more than 60\% of the prefixes belong to the six types we describe in section \ref{sec:ClaInP}, implying that they very likely result from traffic engineering, IP address transfer and failing to aggregate rather than real hijackings. We also find that most of the invalid prefixes are long-lived, which justifies the implication. One possible direction of future work is to do survey over practitioners to verify whether the invalid prefixes are false alarms and find even more types of false alarms.

 \bibliographystyle{IEEEtran}



 \end{document}